\documentclass[%
 aps,
 prb,%
 amsmath,amssymb,
preprint,showpacs,showkeys,%
]{revtex4-1}

\usepackage{graphicx}

\usepackage{amssymb}
 \usepackage{amsthm}

\begin{document}


%
%
%
%
%

\author{V. Romero-Garc\'ia}
 \email{virogar1@mat.upv.es}
 \affiliation{Instituto de Ciencia de Materiales de Madrid, Consejo Superior de Investigaciones Cient\'ificas}
 \altaffiliation{Centro de Tecnolog\'ias F\'isicas: Ac\'ustica, Materiales y Astrof\'isica, Universidad Polit\'ecnica de Valencia.}
 
 \author{A. Krynkin}
\email{A.Krynkin@salford.ac.uk}
\author{O. Umnova}
\email{O.Umnova@salford.ac.uk}
\affiliation{Acoustics Research Centre, The University of Salford, Salford, Greater Manchester, UK.}

\author{L.M. Garcia-Raffi}
\email{lmgarcia@mat.upv.es}
\affiliation{Instituto Universitario de Matem\'atica Pura y Aplicada, Universidad Polit\'ecnica de Valencia.}

\author{J.V. S\'anchez-P\'erez}
\email{jusanc@fis.upv.es}
 \affiliation{Centro de Tecnolog\'ias F\'isicas: Ac\'ustica, Materiales y Astrof\'isica, Universidad Polit\'ecnica de Valencia.}

\date{\today}

\title{Multi-resonant scatterers in Sonic Crystals: Locally Multi-resonant Acoustic Metamaterial}

\begin{abstract}
An acoustic metamaterial made
of a two-dimensional (2D) periodic array of multi-resonant
acoustic scatterers is analysed in this paper. The building blocks consist
of a combination of elastic beams of Low-Density Polyethylene Foam
(LDPF) with cavities of known volume. Sound inside the structure
can excite elastic resonances of the material as well as acoustic
resonances in the cavities producing several attenuation peaks in
the low-frequency range. Due to this behaviour at the subwavelength regime we can define this periodic array as a Locally Multi-Resonant Acoustic Metamaterial (LMRAM) presenting strong dispersive characteristics of the effective properties with subwavelength multi-resonant structural units. The results shown in this paper could be use to design effective Sonic Crystal Acoustic Barriers with wide tunable
attenuation bands in the low-frequency range.

\end{abstract}


\maketitle





\section{Introduction}
\label{sec:Intro}

Artificially designed subwavelength electromagnetic materials,
denoted metamaterials \cite{Veselago67, Smith00, Shelby01}, have motivated in last years a great effort
to develop both theoretically and experimentally the acoustic
analogue metamaterial \cite{Liu00a, Fang06}. Recent works have shown that periodic
distributions of sound scatterers in a fluid, known as Sonic
Crystals (SC)  \cite{Martinez95}, can be employed to design these acoustic
metamaterials. Theoretical predictions
and experimental data have shown in SC the appearance of frequency
ranges, known as Band Gaps (BG), in which sound propagation is not
allowed \cite{Sigalas92, Kushwaha93, Sanchez98}. However, in the regime of large wavelengths, in comparison with the separation between the scatterers, SC behave as effective homogeneous acoustic medium  \cite{Torrent06a}.

The pioneering work of Liu et al.  \cite{Liu00a} proposed a novel
three-dimensional (3D) periodic arrays of coated spheres that exhibited
attenuation bands, whose respective wavelength was about two
orders of magnitude larger than the periodicity of the structure.
The origin of this phenomenon has been explained as due to the
localized resonances associated with each scattering unit. Then,
sound attenuation within the stop bands increases with the number
and density of the local resonators, whereas the resonance
frequency can be tuned by varying their size and geometry. These
results introduce the way towards the acoustic analogous of the
electromagnetic metamaterial, and since
the acoustic wavelength of the attenuation band in the
subwavelenght regime is much longer than the lattice constant of
periodic system, one can define a Locally Resonant Acoustic
Metamaterial (LRAM) whose effective properties can provide an
accurate and simple description of the wave interaction with the
associated LRAM. Since the sound speed in the LRAM is proportional
to $\sqrt{\kappa_{eff} /\rho_{eff}}$, where $\kappa_{eff}$ and
$\rho_{eff}$ are effective modulus and mass density, respectively,
a negative $\kappa_{eff}/\rho_{eff}$ implies an exponential wave
attenuation (evanescent behavior). Although for natural materials
$\kappa$ and $\rho$ are positive numbers to maintain structural
stability, for the effective medium with low-frequency resonances
can appear some differences. It has been shown previously that
low-frequency band gap was induced by an effective mass density
that becomes negative near the resonance frequencies, giving rise
to exponential attenuation of wave. This is in contrast to the
acoustic metamaterials formed by SC with no resonant scatterers
where the effective properties are real and positive.  The
negative $\rho_{eff}$ is seen to result from the coupling of
travelling waves with the local resonances. In particular, the
local resonances enable the strong coupling even though the size
of the locally resonant unit is much smaller than the wavelength
in the matrix, in contrast to the Raleigh scattering case.

\begin{figure*}[hbt]
\begin{center}
\includegraphics[width=145mm,height=80mm,angle=0]{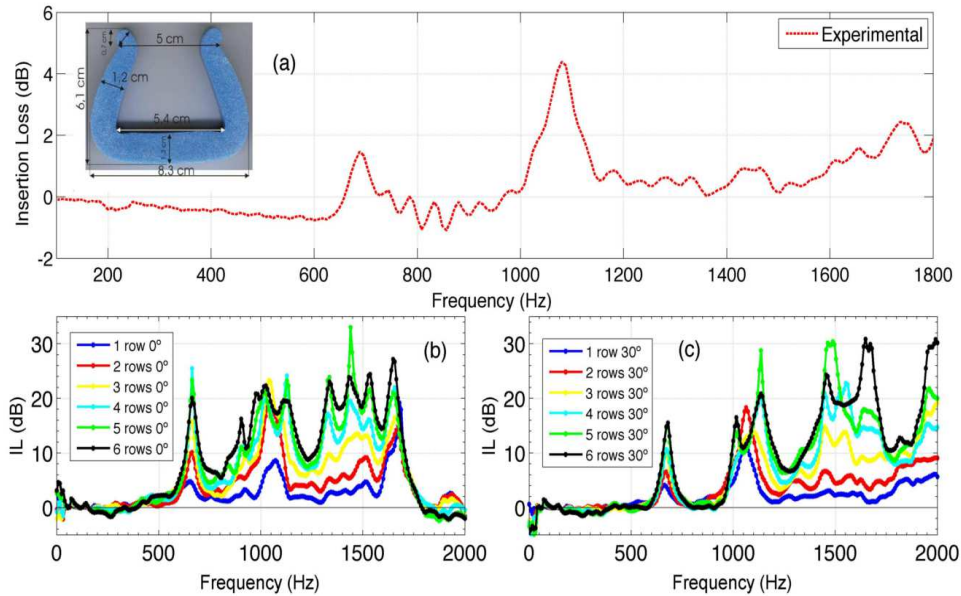}
\caption{Experimental
data. (a) IL of one U-profile. (b) and (c) show the IL of a SC made of different numbers of U-profiles in a triangular array, $a=$12.7 cm, measured at 0$^\circ$ and 30$^\circ$ respectively. The inset shows the transversal view of a U-profile.}
\label{fig:single_scatterer}
\end{center}
\end{figure*}

In this work we present a new kind of acoustic metamaterial made
of a two-dimensional (2D) periodic array of multi-resonant
acoustic scatterers. This Locally Multi-Resonant Acoustic
Metamaterial (LMRAM) have strong dispersive characteristics of the
effective properties with subwavelength multi-resonant structural
units. The building blocks of this acoustic metamaterial consists
of a combination of elastic beams of Low-Density Polyethylene Foam
(LDPF) with cavities of known volume. Sound inside the structure
can excite elastic resonances of the material as well as acoustic
resonances in the cavities producing several attenuation peaks in
the low-frequency range.

The paper is organized as follows. First, in section \ref{sec:pheno}, we
characterize the resonances of a single 2D elastic beam (EB) as well as the resonances of a rectangular cavity. These results allow us to phenomenologically explain the resonances appearing in the scatterers analysed in this work. Because of the complex geometry of the scatterer we numerically solve the acousto-elastic coupled problems. The numerical model is shown in section \ref{sec:num}. Using this model we analyse the behaviour of isolated scatterers as well as the periodic arrays made of them. The experimental validation of all the numerical simulations are shown in section \ref{sec:exp}. Due to the resonant behaviour of the periodic structure in the subwavelength regime we consider this structure as a Locally Multi-resonant Acoustic Metamaterial in section \ref{sec:meta}. Finally, the conclusions of the work are presented in section \ref{sec:conclu}

    \subsection{Motivating results}
        
The scatteres analysed in this work will be called U-profiles due to the geometrical shape of the scatterer (see Figure \ref{fig:single_scatterer}a). In this Section, the interest is focused on the behaviour of the SC made of U-profiles, in the subwavelength regime, meaning that $ka<<\pi$, where $a$ is the periodicity of the array and $k$ is the wavenumber. Hereinafter, this range of frequencies will be called low frequency range.
 
Figure \ref{fig:single_scatterer}a shows the
acoustic response of a commercial scatterer made of the recycled
material: LDPE closed-cell foam (see inset). To do that we have measured the Insertion Loss (IL) defined as the difference between the sound level recorded with and without the sample. The red dashed line in the upper graph illustrates the IL of a U-profile. One can see the existence of two attenuation peaks
appearing in the low frequency range, around 700 Hz and
1100 Hz. These peaks will be called first and second attenuation peaks,
respectively. The nature of both can be understood by
analysing the eigenvalue and scattering problems for the basic
geometrical shapes such as rectangular elastic beam and rectangular cavity, as it will be seen later.

On the other hand, this resonant behaviour could be used to improve
the acoustic behaviour of the SC introducing attenuation peaks in the
range of low frequencies, independent of the incidence direction
of the wave as in the case seen in the previous Section. The attenuation peaks shown in Figure \ref{fig:single_scatterer} are obtained with SC made of U-profiles placed in triangular array with lattice constant $a=12.7$cm.
For this lattice constant it is possible to consider that the upper bound of the low frequency
range corresponds to the first Bragg's frequency of that SC with
value 1545 Hz.

    \section{Phenomenological analysis}
    \label{sec:pheno}
    	The nature of both attenuation peaks below 1500 Hz can be understood by analysing the acoustical properties of basic geometrical shapes like rectangular elastic beams and rectangular resonance cavities. In the next two subsections the resonances of both the elastic beam (elastic resonances) and the rectangular cavity (cavity resonance) are analysed.
	\subsection{Elastic resonances}
\begin{figure*}[hbt]
\begin{center}
\includegraphics[width=150mm,height=70mm,angle=0]{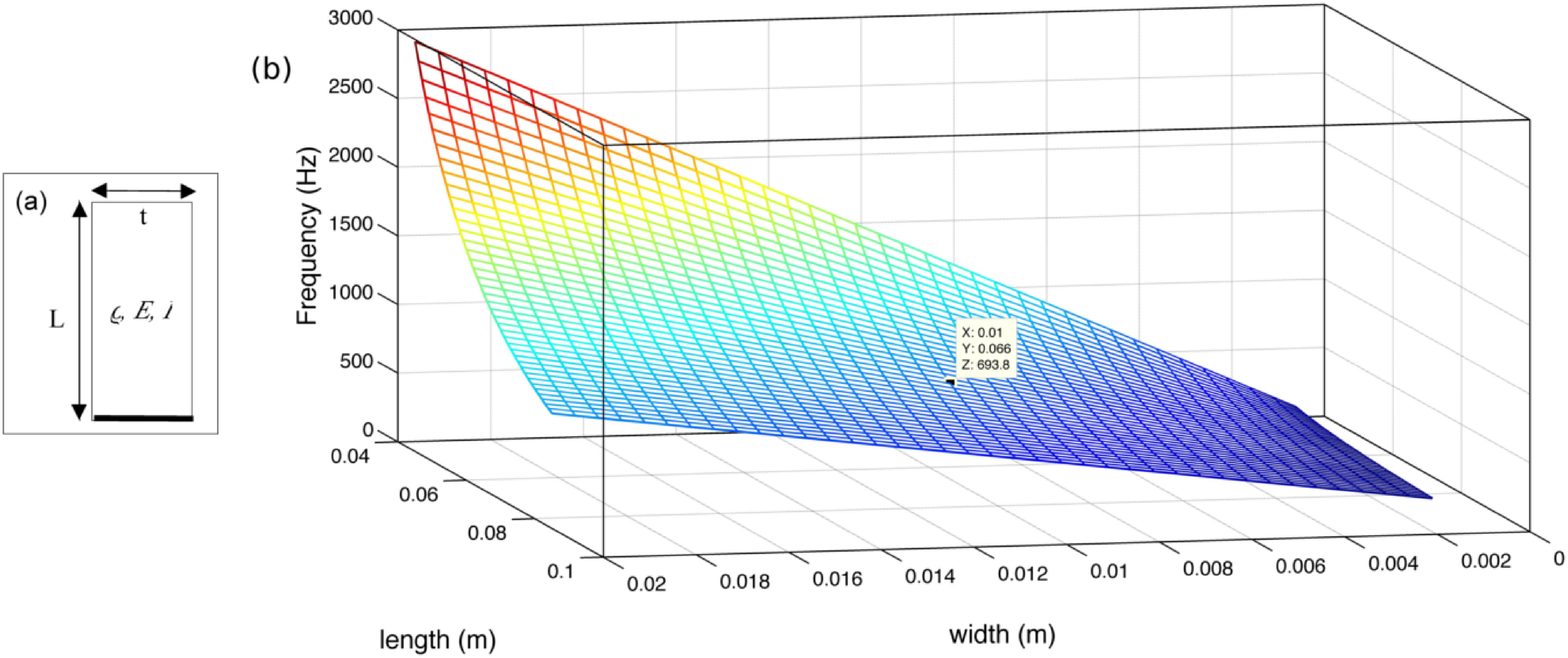}
 \caption{Eigenfrequencies of an
elastic bar of LDPF, density $\rho=$100${kg}/{m^3}$, Young's
modulus $E=$0.35GPa and Poisson's ratio $\nu=$0.4. (a) Schematic
view of the EB. (b) Dependence of the first
eigenfrequency on both the length and the width of the EB made of LDPF.}
\label{fig:dependence}
\end{center}
\end{figure*}

	Consider a 2D elastic beam (EB) made of LDPE foam with
$L$ length and $t$ width (see the schematic view in Figure
\ref{fig:dependence}a); the density of material
$\rho$. Considering that the EB is fixed in one end (black end in Figure \ref{fig:dependence}a), the vibration modes considering temporal harmonic dependence can be analysed by means
of the following Equation  \cite{voltera65, morse68, gere97},
\begin{eqnarray}
E I\frac{\partial^4 v(x)}{\partial
x^4}=-\lambda_m\omega_n^2 v(x)
\label{eq:vibration}
\end{eqnarray}
where $\lambda_m=\rho Lt$ is the linear mass density of the EB,
$E$ is Young's modulus, $\omega_n$ is the
angular frequency of the mode $n$ related to the
frequency as $\omega_n=2\pi\nu_n$ and $I$ is the second moment of
inertia. The $EI$ product is known as flexural rigidity. 

The eigenfrequencies of the EB can be obtained from the
following Equation,
\begin{eqnarray}
\cos(k_n L)\cosh(k_n L)+1=0; \label{eq:modes}
\end{eqnarray}
where $k_n=\sqrt[4]{\omega_n^2\rho L t / E I}$.

From the values of $k_n$ and considering that the material of the
EB is LDPF, the resonance frequencies of the first and second modes
are $\nu_1=693.8$ Hz ($k_1=28.41$ $m^{-1}$) and $\nu_2=4348$ Hz ($k_2=71.12$ $m^{-1}$) respectively. Using Taylor's series, it is possible to approximate
the first mode as
\begin{eqnarray}
k_1\simeq\frac{\sqrt[4]{12}}{L}.
\end{eqnarray}
This first low-frequency solution is particularly interesting for this work. 

In order to validate the use of this analytical approximation, we have compared the results obtained for the first resonance peak with those obtained using COMSOL for several values of thickness-length ratio ($t/L$). These values and the relative differences are shown in Table \ref{tabla}. For the analysed geometry the Euler-Bernoulli equation gives frequency resonance of the elastic beam within 1\% of the COMSOL results.

\begin{table}[hbt]
\begin{center}
\caption{Percentage of the relative differences between analytic
and numerical results
$\left(\frac{\nu_{COMSOL}-\nu_{analytic}}{\nu_{COMSOL}}\times\,100\right)$.}\label{tabla}
\vspace{5mm}
\begin{tabular}{|c|c|c|c|}
 \hline
  & Euler & & Relative \\
 t/L & Bernoulli, & COMSOL, & error \\
  & $\nu$ (Hz) & $\nu$ (Hz) & (\%)\\
 \hline
\hline
 0.01 & 45.7 & 45.8 & 0.22\% \\
\hline
 0.05 & 229 & 228.4 & 0.26\% \\
\hline
 0.1 & 458 & 455 & 0.65\% \\
\hline
 0.15 & 687 & 677 & 1.47\% \\
\hline
 0.2 & 916 & 892.9 & 2.58\% \\
\hline
 0.3 & 1373 & 1300.5 & 5.57\% \\
\hline
 0.5 & 2289 & 1998.7 & 14.52\% \\
\hline
 0.75 & 3434 & 2653.6 & 29.41\% \\
 \hline
\end{tabular}
\end{center}
\end{table}

The resonances of a EB made of a fixed material can be tuned by varying its geometrical
parameters. Figure \ref{fig:dependence}
shows the dependence of eigenfrequency of the first mode on both
the length and the width of the EB made of LDPE foam. The black point marks the position of the first resonant frequency for the geometrical properties of the EB analysed in this Section ($t=0.01$ m and $L=0.066$ m).
One can observe that the bigger the length, the lower the frequency
of the first mode, and that the bigger the width, the higher the
frequency of the first mode. The intersection point of the black lines
corresponds to the eigenfrequency for the EB with the geometrical
parameters considered in this work. It can be also observed that the eigenfrequencies grow linearly with the width of the EB, however the growth with the length is not linear.

\subsection{Cavity resonances}
Another interesting property of the U-profiles is that they present,
in addition to the elastic properties of the material, a cavity where
sound could be localized due to resonances. Several works in
the literature have analysed the effect of cavity resonators in
periodic structures. In addition to the BG of the periodicity, the systems made of resonators show low frequency attenuation bands produced by the resonances of
Helmholtz or split-ring resonators.
Due to both 
the control of the resonances of the EB and the resonance of the
cavity, periodic structures made of U-profiles elastic scatterers
can be easily tunable in the range of low frequencies.

\begin{figure}[hbt]
\begin{center}
\includegraphics[width=80mm,height=40mm,angle=0]{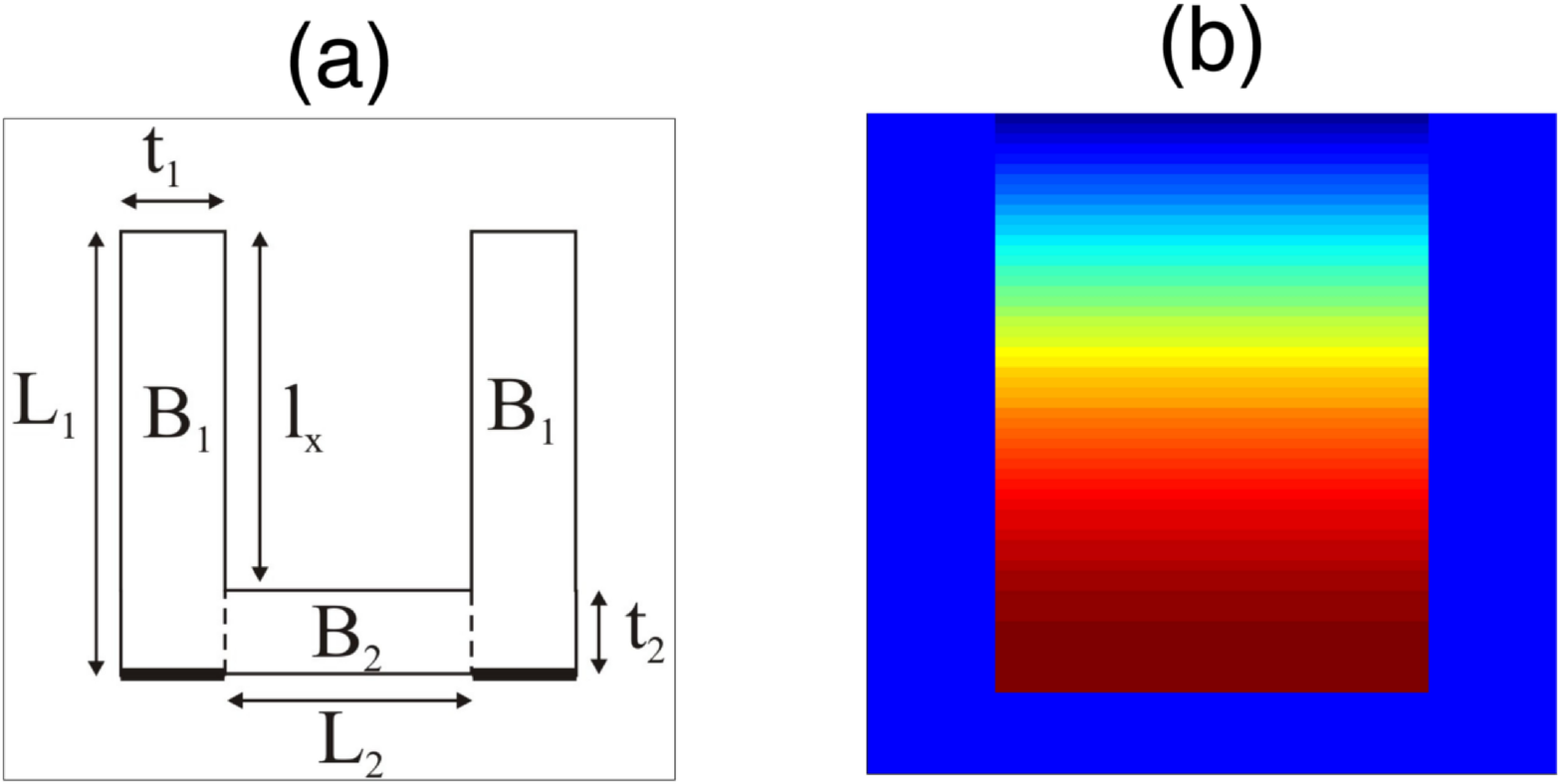}
\caption{U-profile elastic scatterers. calculated using FEM. (a)
Geometrical shape and parameters characterizing the size of
the U-profile. (b) Acoustic field inside the cavity for the resonant
frequency $\nu_r$.}
\label{fig:U_profile} 
\end{center}
\end{figure}

	The eigenfrequencies of a rectangular cavity with several
boundary conditions has been widely analysed in the literature. As one
can see in Figure \ref{fig:U_profile}a, the cavity of the
U-profile has a length and a width equal to $l_x=0.066$ m and $L_2=0.04$ m
respectively. To solve the problem, one can consider that the walls of
the U-profile are perfectly rigid. Thus, Neumann boundary
conditions should be considered at the boundaries, and Dirichlet conditions in the boundary in the open side of the U-profile.
The solution of the analytical problem results in an
eigenfrequency problem whose fundamental mode has a eigenfrequency equal to
$\nu_{r}=c_{air}/(4l_x)$. However, we have shown that the air
immediately outside the end of the cavity takes part in the
acoustic oscillation. This air makes the cavity appear to be
acoustically somewhat longer than its physical length. This effective length gives rise to a shift of the resonance frequency, for this reason it is called in the literature, the end correction of the cavity. In order to compute
the correct resonance frequency, this effective length and the corresponding frequency correction have to be considered. A more rigorous
analysis of the cavity would be required to find the exact
resonance frequencies, but it has been assumed that the resonance is affected by the end correction of the cavity of the
U-profile following the next equation \cite{Norris05}
\begin{eqnarray}
\nu_{r}=\frac{c_{air}}{4(l_x+\delta)},
\end{eqnarray}
where
\begin{eqnarray}
\delta\simeq\frac{4\sqrt{\frac{l_xL_2}{\pi}}\arcsin{\left(\frac{L_2}{2a_e}\right)}}{\pi}\log{\left(\frac{1}{\arcsin{\left(\frac{L_2}{2a_e}\right)}}\right)}=0.0076.
\end{eqnarray}
is the end correction used in \cite{Norris05}. The adaptation of the equation C29 in reference \cite{Norris05} can be seen in the \ref{ap1}.
Then, in the case of the cavity considered in this work, the
frequency of the first mode is $\nu_r=$1155 Hz. 

    \section{Numerical results}
    \label{sec:num}
        \subsection{FEM model}
    \label{sec:FEM}
We start this Section by analysing the propagation of acoustic waves inside periodic structures made of
solid scatterers, $B$, embedded in a fluid host, $A$, using FEM. Due to the physical properties of the host
material, the eigenmodes of the whole system are pure longitudinal
waves, while transverse modes cannot propagate. Then, the governing
Equation in $A$ is
\begin{eqnarray}
-\frac{\omega^2}{c_A^2}p=\nabla\left(\frac{1}{\rho_A}\nabla
p\right)
\end{eqnarray}
where $p$ is the pressure, $\rho_A$ is the density and $c_A$ is
the sound velocity in the host material.

The propagation of elastic waves inside the scatterers, locally
isotropic medium, is governed by
\begin{eqnarray}
-\rho_B\omega^2u_i=\left\{\frac{\partial \sigma_{ij}}{\partial
x_j}\right\}, \label{eq:acoustic}
\end{eqnarray}
where $\rho_B$ is the density of the elastic material and $u_i$ is
the $i$th component of the displacement vector. The stress tensor
is defined by
\begin{eqnarray}
\sigma_{ij} & = &\lambda_Bu_{ll}\delta_{ij}+2\mu_Bu_{ij}\nonumber\\
u_{ij} & = & \frac{1}{2}\left\{\frac{\partial u_i}{\partial
x_j}+\frac{\partial u_j}{\partial x_i}\right\},
\label{eq:elastic}
\end{eqnarray}
where $\lambda_B$ and $\mu_B$ are the Lam\'e coefficients.

In this problem the acoustic wave is incident on the scatterer and then
the pressure acts as a load on the elastic medium. On the other
hand, the elastic waves in the scatterer act as an additional acceleration on the acoustic field. In order to simultaneously solve
\ref{eq:acoustic} and \ref{eq:elastic} we introduce
the following boundary conditions,
\begin{eqnarray}
\frac{\partial p}{\partial n}|_{\partial B} & = &
\rho_A\omega^2\vec{u}\vec{n}\\
\sigma_{ij}n_j|_{\partial B} & = & -pn_i\nonumber.
\end{eqnarray}
where $\partial B$ is the boundary of the medium $B$ and $\vec{n}$
is the outward-pointing unit normal vector seen from inside the
scatterer medium.

To solve the stated problem COMSOL MULTIPHYSICS has been used, as well as a finite-element analysis and solving software package. FEM is a good technique when the shape of the involved objects is complicated and several physical problems are coupled. In the numerical problem, the domain in which the solution is obtained was surrounded by Perfectly Matched Layers (PML) region in order to emulate
the Sommerfeld radiation condition in the numerical solution \cite{Berenguer94}. 


    \subsection{Scattering problem}
    
    \subsubsection{Single scatterer}
\begin{figure}[hbt]
\begin{center}
\includegraphics[width=90mm,height=50mm,angle=0]{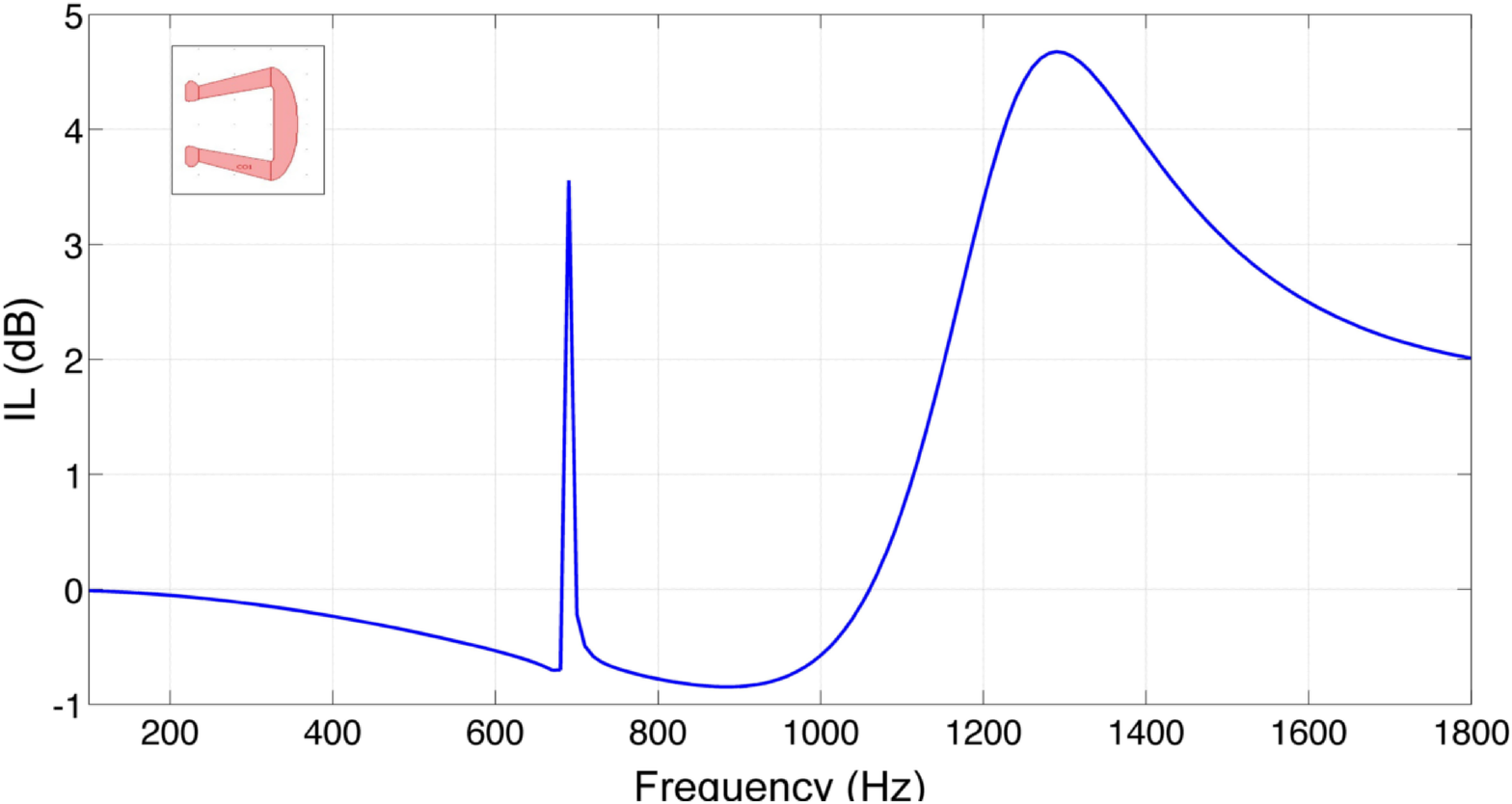}
\caption{Numerical results of single scatterer. IL produced by a U-profile. The inset shows a image of the numerically modelled U-profile. }
\label{fig:Single_scatterer_num}
\end{center}
\end{figure}

    First of all, the frequency response of the U-profile will be analysed. The geometry of the U-profiles was implemented using the CAD tools of  COMSOL. In the inset of Figure \ref{fig:Single_scatterer_num}, one can see the considered model of the scatterer, this geometry slightly differs from the real one. In the numerical model, a plane wave impinging the scatterer from the left side has been considered and the IL at a point behind the scatterer is calculated.
    
   Figure \ref{fig:Single_scatterer_num} shows the numerical results obtained using COMSOL. One can observe similar frequency response to the one experimentally observed in Figure \ref{fig:single_scatterer}. Two attenuation peaks numerically calculated appear near 700 Hz and 1200 Hz. A sensibility analysis varying the geometry of the U-profile was done, and one can observed that the first peak is very sensitive to changes in geometry. However the second one does not substantially change with the variations of the geometry of the U-profile. These changes are in good agreement with the predictions of the resonances of a rectangular elastic beam profile with a resonant cavity. 
    
    As we have previously explained, the first peak corresponds to the resonances of the EB of the U-profile, therefore small changes in the geometry can produce high changes in the resonant frequencies  (see Figure \ref{fig:dependence}). This provides a powerful design tool in the first of the low frequency attenuation peaks.
    
    On the other hand, if we compare Figures \ref{fig:single_scatterer} and \ref{fig:Single_scatterer_num} we can observe again a difference between the numerical and experimental frequencies of the second peak. The modelling of wave scattering in FEM should be able to describe any effects within the theory of linear acoustic and solid mechanics. Use of the end correction could be only appropriate for analytical approximation which neglect the interactions between the cavity interior and the environment. Any discrepancies observed between numerical results and experiments are most likely due to the idealization of the scatterer material. 
    
     \subsection{Infinite periodic array of U-profiles}
\begin{figure*}[hbt]
\begin{center}
\includegraphics[width=130mm,height=70mm,angle=0]{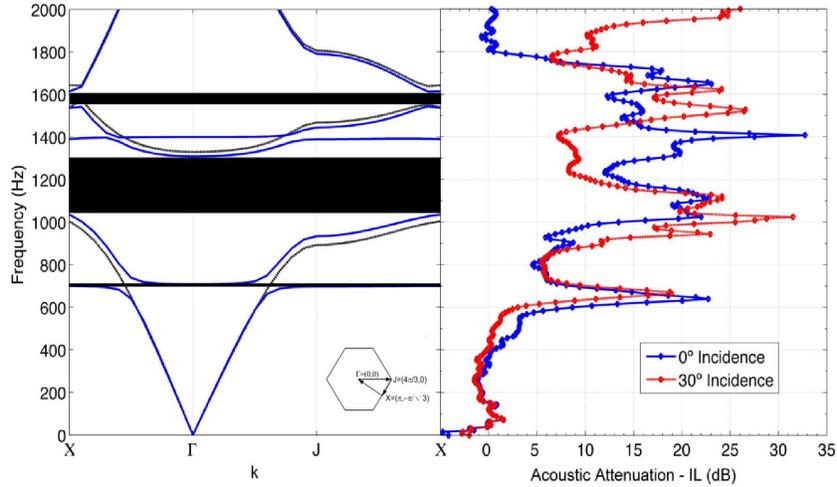}
\caption{Band Structures for a periodic arrangement of U-profiles in triangular lattice of $a=$12.7 cm. Left panel: Black dashed line represents the band structures for a rigid U-profiles, whereas the blue line represents the bands for the elastic U-profile. The black surfaces indicate the non propagating ranges of frequencies. The Brillouin zone is defined as $\Gamma=(0, 0)$, $X=(\pi, -\pi/\sqrt{3})$, $J=(4\pi/3, 0)$. Right panel: Measured IL of a triangular lattice of elastic U-profiles measured in the two main symmetry directions, 0$^\circ$ (blue line) and 30$^\circ$ (red dahsed line) }
\label{fig:Triangular_blue_bands}
\end{center}
\end{figure*}

    We analyse the propagating properties of a periodic arrangement of U-profiles by means of the application of the Bloch periodic boundary conditions in a unit cell. That means studying the band structures for a periodic array of U-profiles.
    Figure \ref{fig:Triangular_blue_bands} shows the Band structures of a periodic arrangement of U-profiles arranged in triangular lattice of $a=$12.7 cm. The black line represents the band structures considering the perfectly rigid U-profiles, this means that we can consider Neumann boundary conditions in the wall of the scatterer. One can observe that this arrangement presents the pseudogaps related to the periodicity as well as the stop band due to the resonance of the cavity.
    
    If the elastic properties of the U-profiles are considered, then the band structures are represented by the blue continuous line. One can observe the BG due to the periodicity ($\simeq$1600 Hz), the stop band of the resonance of the cavity (1100 Hz) and the resonance of the elastic beams of the U-profile (700 Hz). In the representation the non propagating ranges of frequencies are presented by the black surfaces.
    
In order to compare the numerical results with those experimentally obtained a new plot was added in the right panel, where the IL of a triangular periodic distribution of U-proflies was measured in the main symmetry directions: 0$^\circ$ (blue line) and 30$^\circ$ (red line). One can observe the low dependence on the direction of incidence of the resonance due to the vibration of the elastic beams and the resonance of the cavity, but the directionality appears in the frequencies of the BG due to the periodicity.
    
      \subsubsection{Finite periodic array of U-profiles}
    Once the numerical results of the acoustical behaviour of an isolated U-profile have been analysed, the next step is to analyse a periodic distribution of this elastic scatterers following a triangular lattice with lattice constant $a=0.127$ m. Here, a plane wave impinging from the left side is considered and the numerical domain is again surrounded by a PML region. Thus the numerical solution accomplishes an approximated Sommerfeld condition. 
    
\begin{figure}[hbt]
\begin{center}

\includegraphics[width=85mm,height=50mm,angle=0]{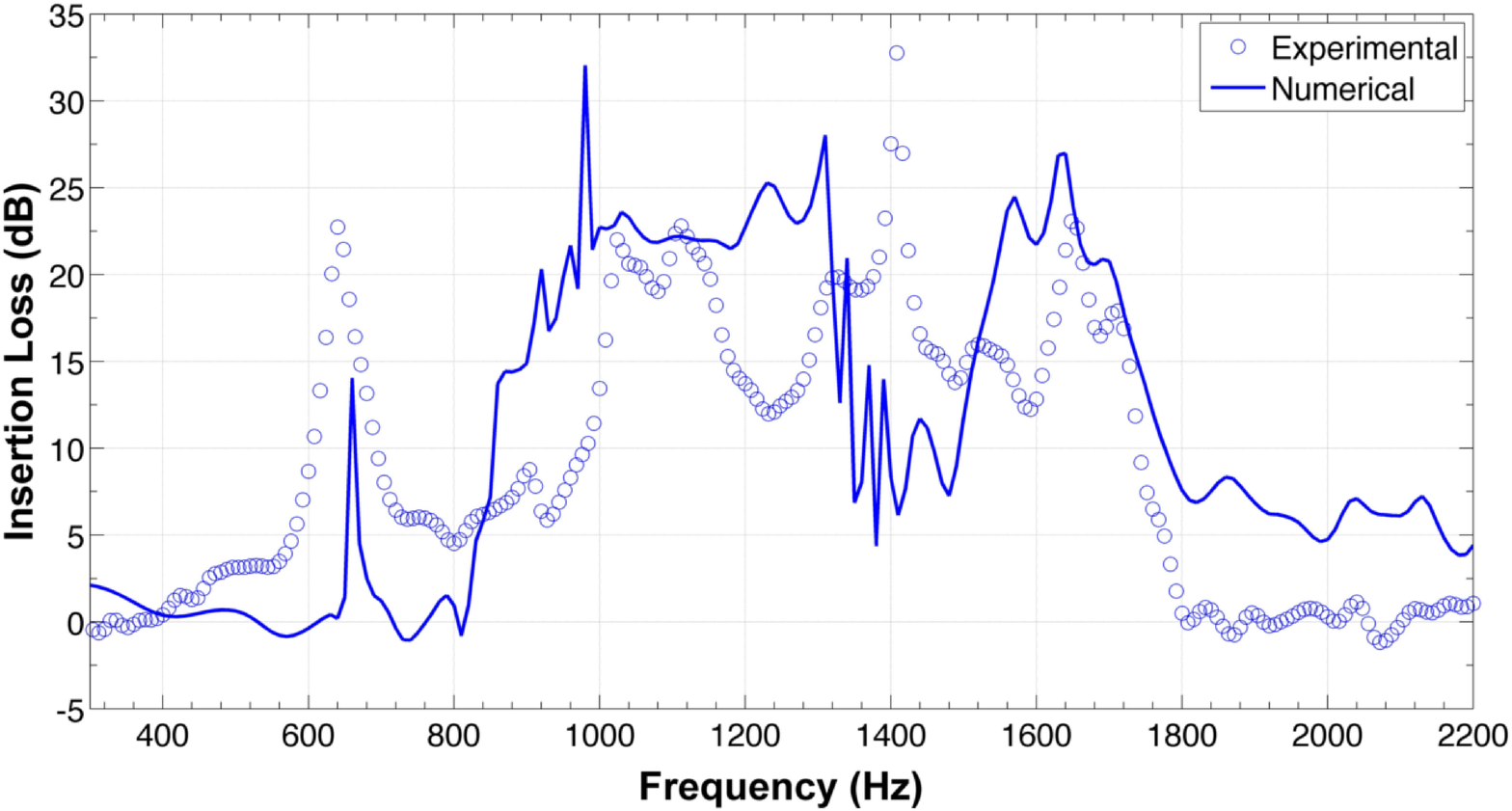}

\caption{IL of an array of U-profiles. Blue continuous line represents the numerically predicted IL and open circles represent the measured IL. A plane wave impinges the structure from the left side to the right side. }
\label{fig:numerical_array}
\end{center}
\end{figure}

    The blue continuous line in Figure \ref{fig:numerical_array} shows the numerically predicted IL of a finite structure made of 6 rows of 10 U-profiles for an incident wave in the direction of 0$^\circ$. By comparing the results of the scattering of an isolated U-profiles (see Figure \ref{fig:single_scatterer}), one can observe that the resonance of the elastic beams, as well as the cavity resonance, have been increased due to the increase in the number of resonators. Moreover, an attenuation peak around 1600 Hz appears and it can be related to band gap of the array.
    
    Open circles in Figure \ref{fig:numerical_array} shows the measured IL for the same array as the numerically calculated is also plotted. One can observe a good correspondence between the attenuation peaks numerically predicted with those experimentally obtained. However, the experimental attenuation peak related to the elastic resonances presents a higher attenuation level than the one numerically predicted. A possible explanation for this effect could be the existence of some absorption effect of the material that it is not considered in the model.
    
    Then, the resonance effect of the scatterers is not destroyed by the multiple scattering inside the structure, therefore one can combine resonances with  multiple scattering in order to obtain several attenuation peaks. As we explained in the beginning of this Section, an interesting feature of these scatterers is that they show two resonances in the range of frequencies below the first BG of the periodic structure.
    
    \section{Experimental results}
    \label{sec:exp}
    The acoustic attenuation
capabilities of the single LDPE foam scatterers as well as of the periodic
structures of these scatterers have been measured in terms of
IL. The sample is excited by white noise.

\subsection{Single scatterer}
The experimental analysis of the single LDPF scatterers has been
divided into two parts, acoustic and vibrational analysis.

\begin{figure*}[hbt]
\begin{center}
\includegraphics[width=135mm,height=85mm,angle=0]{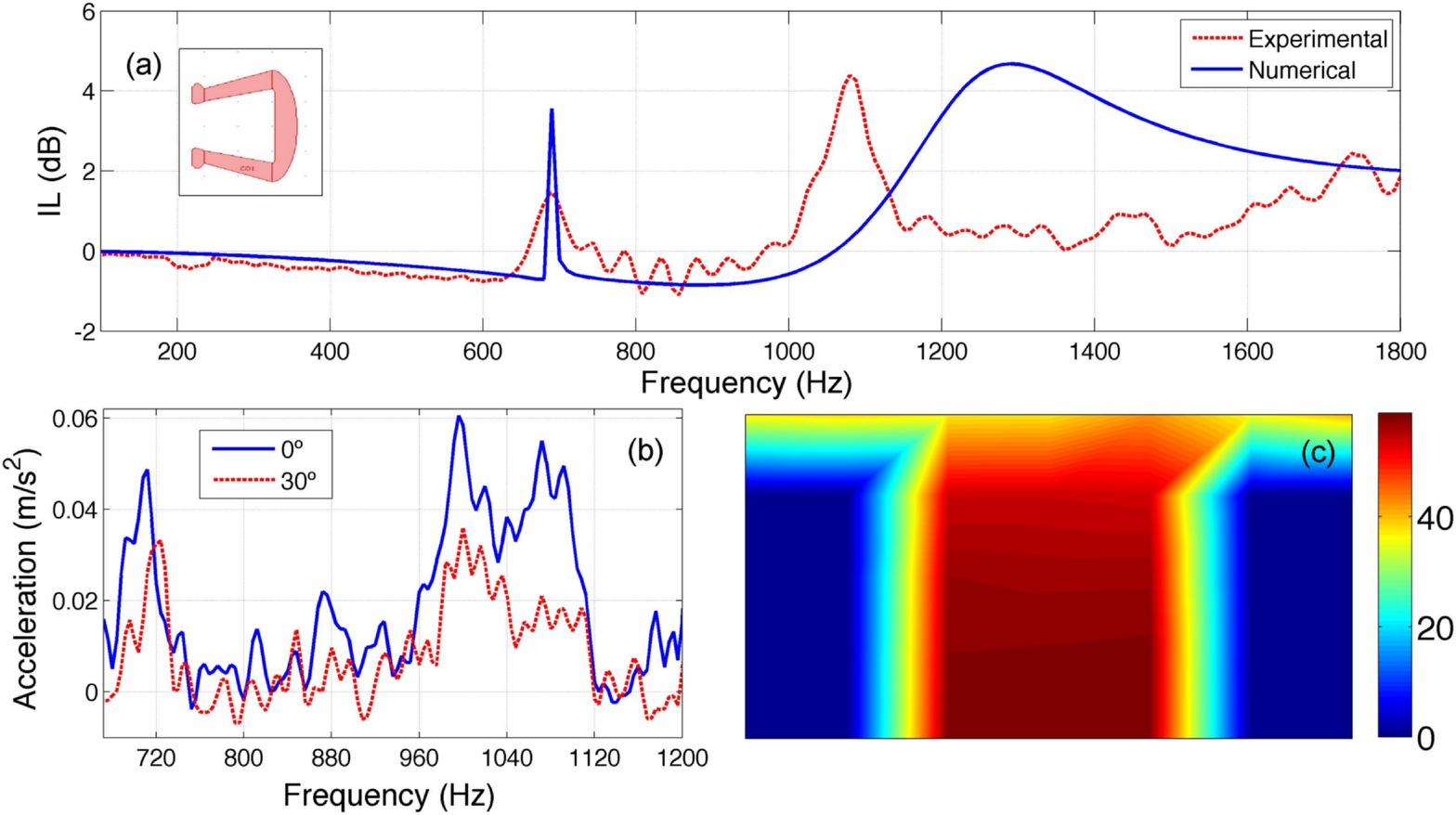}
\caption{Experimental
results of a single scatterer. (a) IL (dB) measured behind the
scatterer. Red dashed line represents the experimental results and
Blue line represents the numerical simulation using FEM, (b)
Experimental measurements of the vibration of the EB of the LDPE foam
scatterer. Blue line represents 0$^\circ$ of incidence and red dashed
line represents 30$^\circ$ of incidence. (c) Sound level map measured
inside the cavity for the resonant frequency for $\nu=$1104 Hz.
Step $\Delta x=\Delta y=$1 cm.}
\label{fig:Experimental_single_scatterer} 
\end{center}
\end{figure*}

The IL of the single scatterer is shown in Figure
\ref{fig:Experimental_single_scatterer}a in red dashed line. The real shape of the scatterer was modelled using FEM. Using the
acoustic-elastic coupling previously presented, the IL was numerically obtained (see the blue line in Figure
\ref{fig:Experimental_single_scatterer}a). One can observe that the
numerical method previously used is in good agreement with the
experimental results. The observed discrepancy between the numerical
and experimental results in the second peak can be explained by the lack of precision in the representation of the scatterer profile in the numerical method.

In Figure \ref{fig:Experimental_single_scatterer}a one can see the
presence of the cavity and the EB resonances. In order to experimentally explain 
these resonances, we have measured on the one hand the vibration of
the EB with an accelerometer, and on the other hand, the acoustic
field inside the cavity of the LDPE foam for the resonant frequency of
the cavity. In Figure \ref{fig:Experimental_single_scatterer}b, one
can observe the vibration of the EB for two different incidence
direction of the acoustic wave. Blue continuous line represents the vibrations of the wall of the U-profile for the incident wave in the $\Gamma$X direction. Red dashed line represents the vibrations of the walls for a wave impinging in the $\Gamma$J direction. The vibration of the wall of the elastic beam
increases at the resonant frequencies.  Figure \ref{fig:Experimental_single_scatterer}b shows the increasing in the vibration of the wall in the resonance of the material and in the cavity resonance independently of the
incident direction of the acoustic wave. 

Figure
\ref{fig:Experimental_single_scatterer}c shows the
acoustic field inside the cavity of the LDPE foam obtained by moving
the microphone with our robotized acoustic measurement system in $1$ cm steps inside the cavity.
The field inside the cavity is similar to the one numerically obtained in Figure \ref{fig:U_profile}d. The resonance of the
cavity induces the vibration of the walls as observed in Figure
\ref{fig:Experimental_single_scatterer}b.

\subsection{Periodic array}

\subsubsection{Dependence on the number of resonators}
\begin{figure*}[hbt]
\begin{center}
\includegraphics[width=65mm,height=40mm,angle=0]{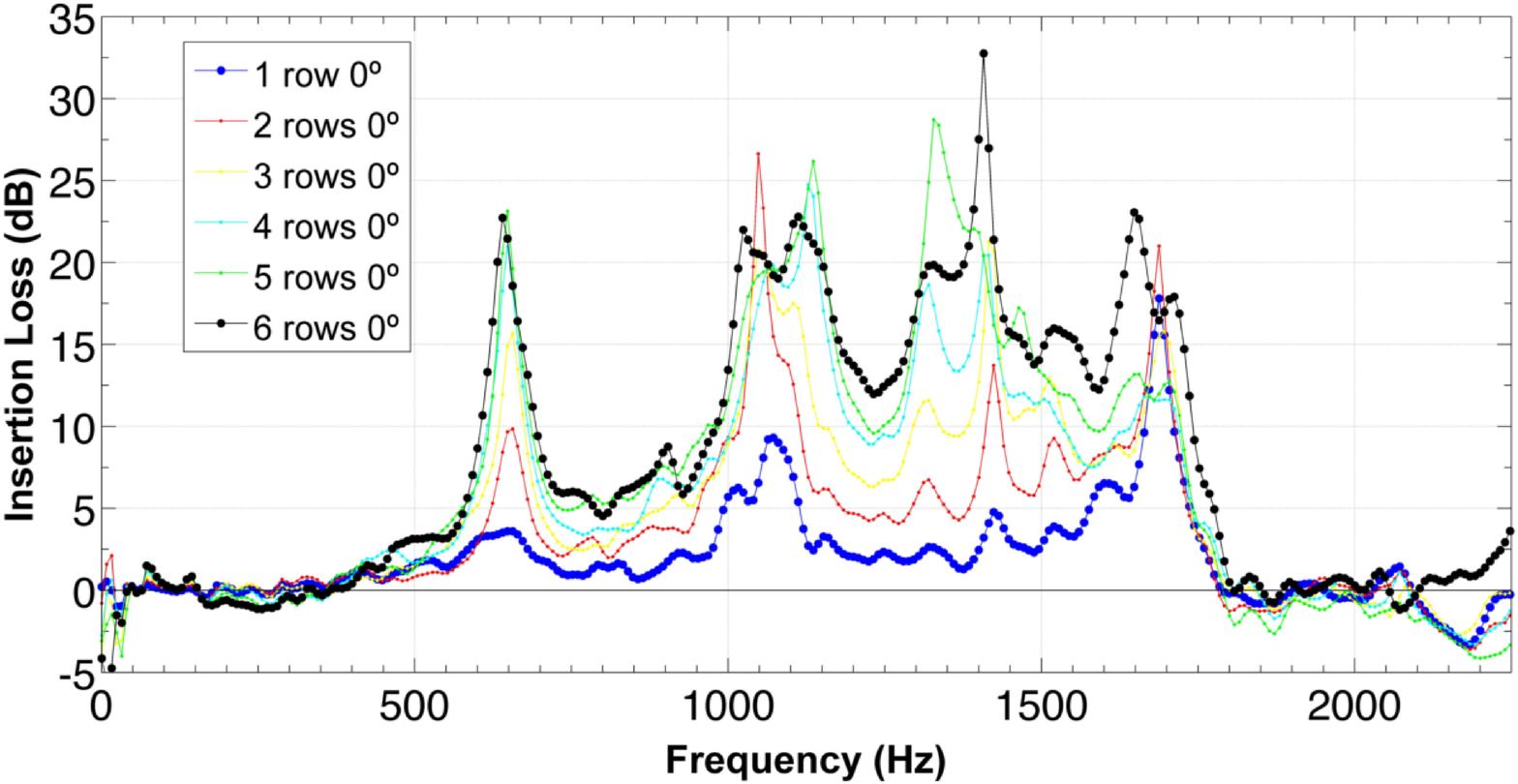}
\includegraphics[width=65mm,height=40mm,angle=0]{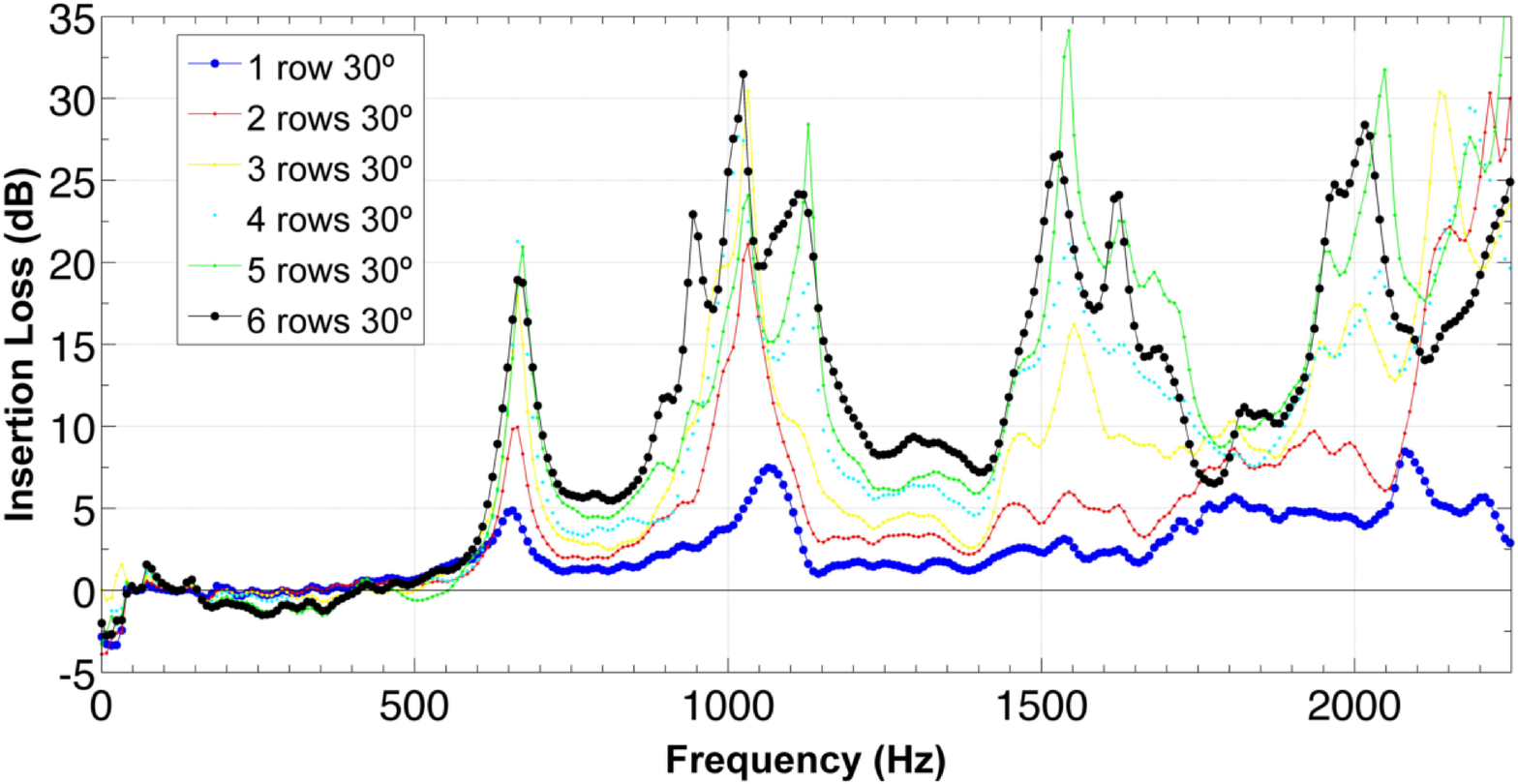}\\
(a)\hspace{6cm}(b)
\caption{Experimental measurement of the IL for determining the dependence of the attenuation peaks on the number of scatterers. Open colored circles represent the IL for six structures made of different number of rows (from 1 to 6 rows of 10 cylinders per row). IL measured 1 m away from the end of the complete structure. (a) Measurement in the $\Gamma$X direction. (b) Measurements in the $\Gamma$J direction. }
\label{fig:num_res} 
\end{center}
\end{figure*}

One can expect that both the resonant effect and the multiple scattering depend on the number of scatterers in the array. To prove it, we built and measured six configurations with an increasing number of scatterers. The final structure presents 6 rows of 10 scatterers per row. We measured the IL at the same point for the six structures that have from 1 to 6 rows respectively. Figure \ref{fig:num_res} shows these experimental results. The coloured open circles represent the IL measured 1 m away from the end of the complete crystal. Blue open circles show the IL of a structure made of 1 row of U-profiles for 0$^\circ$ incidence whereas the black open circles represent the IL of structure made of 6 rows of U-profiles for 30$^\circ$ of incidence.

We can observe that both resonances due to the elastic material and the cavity, depend on the number of cylinders in the structure. Also, in the case of only 1 row, where there is no periodicity in 2D, the resonance peaks are present in the attenuation spectrum whereas Bragg's peaks do not appear. The attenuation spectra of structures made of rigid scatterers always present ranges of frequencies where there is sound reinforcement, meaning negative IL. However, it is interesting to note that these structures do not present ranges of frequencies with sound reinforcement.

\subsubsection{Dependence on the incidence direction}
One of the main characteristics of the stop bands produced by periodic arrays is their dependence on the incident direction. In periodic systems the BG results from the intersection of the frequency ranges of the pseudogaps in the main symmetry directions, the upper and lower bands of each main direction being dependent on the incident direction. However, it is known that the resonance effect must be independent from the incidence direction.

\begin{figure}[hbt]
\begin{center}
\includegraphics[width=90mm,height=50mm,angle=0]{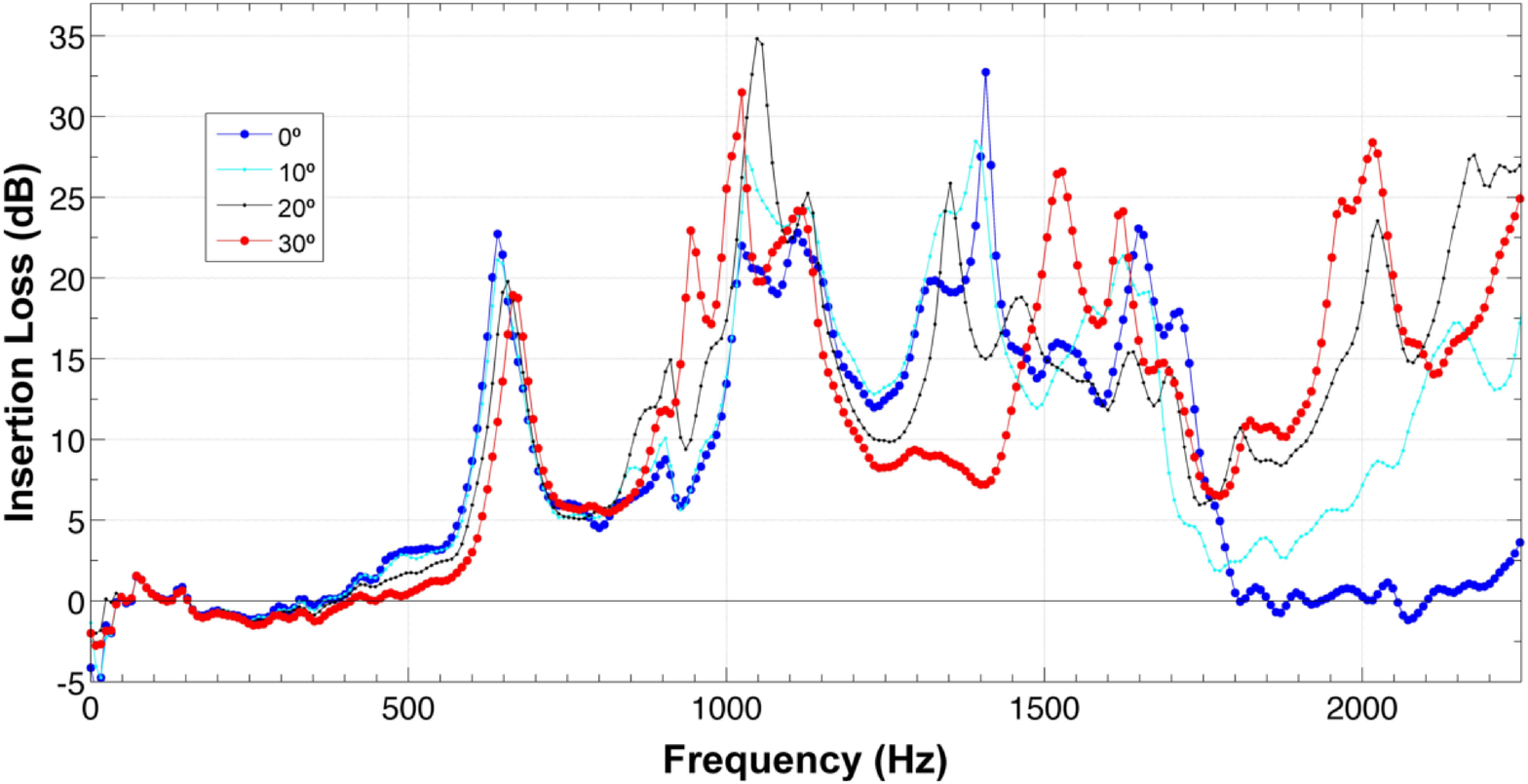}
 \caption{Experimental measurement of the IL to determine the dependence of the attenuation peaks on the direction of incidence of the wave. Open coloured circles represent the IL for four different directions, 0$^\circ$, 10$^\circ$, 20$^\circ$ and 30$^\circ$. IL measured 1m away from the end of the complete structure (6 rows of 10 cylinders per row).}
\label{fig:directionality}
\end{center}
\end{figure}

Here, we measure this dependence of both  the resonance and the multiple scattering in a periodic array of U-profiles in the direction of incidence. We have especially measured the IL of a complete structure for several incident directions, between the two main symmetry directions (0$^\circ$ and 30$^\circ$). Figure \ref{fig:directionality} shows these experimental results measured in the anechoic chamber.

One can observe in Figure \ref{fig:directionality} the low dependence on incidence direction of the attenuation peaks produced by the resonances of the elastic walls and by the cavity of the U-profile. However, one can see that the behaviour of the attenuation peak produced by the multiple scattering in the periodic system is highly dependent on the incidence direction.

    \section{Locally multi-resonant acoustic metamaterial}
    \label{sec:meta}

In LRAM, the sound speed is proportional to $\sqrt{\kappa_{eff} /\rho_{eff}}$, where $\kappa_{eff}$ and
$\rho_{eff}$ are the effective modulus and the mass density, respectively. 
Depending on the values of these parameters the metamaterial presents several responses in frequency. SC made of rigid scatterers with no resonant properties can be analysed as an acoustic metamaterials showing real and positive effective properties  \cite{Torrent07}. However, some interesting differences can appear due to the effective medium with low-frequency resonances.

In order to have a propagating plane wave inside the medium, we should have either both positive $\kappa_{eff}$ and $\rho_{eff}$ or both negative $\kappa_{eff}$ and $\rho_{eff}$. Moreover, with these values the Poynting vector for a propagating plane wave is defined by
\begin{eqnarray}
 \vec{S}=\frac{\imath}{2\omega \rho}p\nabla p*=\frac{|\vec{p}|^2\vec{k}}{2\omega \rho}.
 \end{eqnarray}
If $\kappa_{eff}$ and $\rho_{eff}$ are positive, the Poynting vector, $\vec{S}$, presents the same direction as $\vec{k}$ and the Snell law is normally accomplished. However if $\kappa_{eff}$ and $\rho_{eff}$ are negative, $\vec{S}$ and $\vec{k}$ present opposite directions, and the metamaterial behaves as a left handed material, where the negative refraction appears. Physically, the negativity of $\kappa_{eff}$ and
$\rho_{eff}$ means that the medium displays an anomalous response at some frequencies such that it expands upon compression (negative bulk modulus) and moves to the left when being pushed to the right (negative density) at the same time. These materials present unique properties due to the double-negative medium, such as negative refractive index and subwavelength focusing  \cite{Guenneau07}.

However, if only one of both parameters $\kappa_{eff}$ and
$\rho_{eff}$ are negative, the sound velocity is complex, and the vector   presents a complex value. Thus, when the real component of the expression of the Poynting vector is negative and sufficiently large, we can observe a narrow frequency range, corresponding to the region of negative modulus, where $Re(\vec{k}\vec{S}) < 0$. A direct consequence of such behaviour is the exponential wave attenuation in such frequencies. It has previously been shown that
low-frequency attenuation bands can be induced by an effective bulk modulus that becomes negative near the resonance frequencies, giving rise to exponential decay of modes  \cite{Fang06}.

In the system studied here, the stop bands at low frequencies are independent of
the angle of incidence and of the lattice constant\footnote{We have also experimentally observed non dependence on the height of the U-profiles.}. Moreover, there is not transmission wave in the resonance frequencies, consequently we do not observe any negative refraction or subwalength imaging near the resonance. These properties would imply propagation in some frequency region. Thus, we can conclude that, as in the case of reference \cite{Fang06}, the periodic structure made of U-profile could be represented by one of the effective parameters being possitive and the other one negative in the resonant frequencies.

  A rigorous parameter retrieval procedure on
the line of those developed for electromagnetic and acoustic
cases will be required to be implemented on this system to obtain
the $\kappa_{eff}$ and $\rho_{eff}$. But since the LRAM structure,
in our case, is not a far subwavelength in size of the operating
frequency ($\lambda \sim a$), such a homogenization of all
properties via effective medium parameters is difficult. However,
one can follow the formalism of the electromagnetic material to
phenomenologically analyse the behaviour of the system in the
subwavelength regime analogously as in reference  \cite{Fang06}.

Up to the best of our knowledge, the negativity of both the bulk modulus and the effective density in acoustic  metamaterials is related to the monopolar and the dipolar resonances of the building blocks of the metamaterial respectively. The monopolar Helmholtz \cite{Hu08} resonances have been used to design acoustic metamaterials with negative bulk modulus and the dipolar resonances of spheres have been used to design double-negative metamaterials \cite{Li04}, i.e., negative bulk modulus and dynamic mass density. In this work we have shown evidence about the different origin of the resonances in the U-profile system being the two attenuation peaks produced by monopolar resonances. Therefore both resonances could be considered in the expression of the bulk modulus.

\begin{figure}[hbt]
\begin{center}
\includegraphics[width=110mm,height=55mm,angle=0]{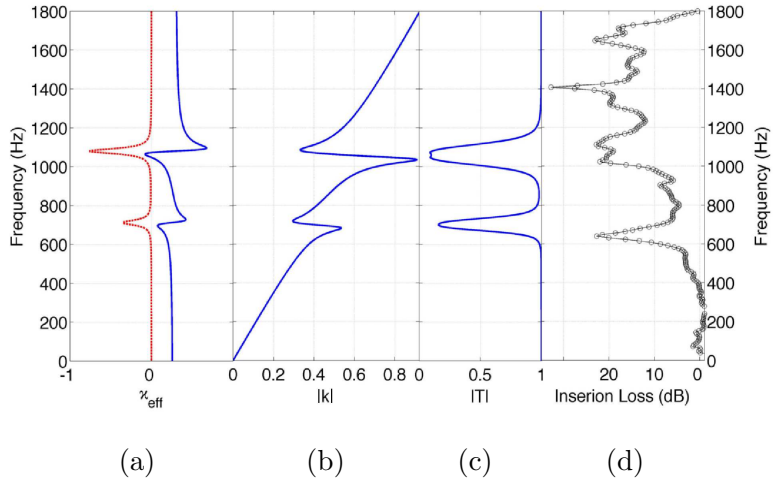}\\
(a)\hspace{2cm}(b)\hspace{1.5cm}(c)\hspace{1.5cm}(d)
\end{center}
\caption{Effective parameters. (a) Effective bulk
modulus. Imaginary part is plotted in red dashed line whereas real
part is plotted in blue continuous line. (b) Dispersion relation.
(c) Transmission coefficient for a slab of metamaterial with
$L_{eff}=$0.549 m. (d) Measured IL of an array of U-profiles.}
\label{fig:keff}
\end{figure}

The acoustic properties of a 2D SC can be mapped into an
electromagnetic counterpart, where $p$, $\vec{v}$, $\rho$, $k$
correspond to $H_z$, $\vec{E}$, $\varepsilon$, $\mu$,
respectively. Following the formalism of the electromagnetic
metamaterials, one can consider that the systems behave as a
metamaterial with an effective bulk modulus $\kappa_{eff}(\omega)$ in
the form,
\begin{eqnarray}
\label{eq:effective_modulus}
\kappa_{eff}^{-1}(\omega)=\sum_{j=1}^{N_{res}}\left(\frac{E}{3(1-2\nu)}\right)^{-1}\left(1-\frac{F\omega_{0j}^2}{\omega^2-\omega_{0j}^2+\imath\Gamma\omega}\right),
\end{eqnarray}
where $F$ is the filling fraction, $\omega_{0j}$ represents the
resonant frequencies of the LDPF scatterer, $\Gamma$ is the
dissipation loss in the resonating elements and $N_{res}$ is the
number of resonances of the scatterers. The loss term $\Gamma$
cannot be determined a priori therefore it should be experimentally determined. In our system $N_{res}=2$ and the
resonances are represented by:
\begin{eqnarray}
\omega_{01}=\frac{\sqrt{12}}{L^2}\left(\frac{\rho L t
}{EI}\right)^{-1/2},\\
\omega_{02}=\frac{2\pi c_{air}}{4(l_x+\delta)}.
\end{eqnarray}
In Figure \ref{fig:keff}a, one can observe the effective bulk
modulus of the material. The imaginary part of the effective bulk modulus
presents this particular frequency-dependent
response which is essential to the range of
frequencies where a stop band is expected.

To obtain the transmission coefficient of a
slab of the metamaterial with the bulk modulus shown in Equation
\ref{eq:effective_modulus}, it is necessary to determine the size
of the effective thickness of the material. The filling fraction of the structure
is,
\begin{eqnarray}
f=\frac{\sum_{i=1}^{N}A_{cyl}}{A_{eff}}
\end{eqnarray}
where $N$ is the number of scatterers, $A_{cyl}$ is the area of each
scatterer and $A_{eff}$ is the area occupied by the homogeneous
scatterer. For a homogeneous scatterer of $N$ scatterers of area
$A_{cyl}$ in a lattice whose unit cell has an area $A_{uc}$, the filling
fraction gives the following Equation
\begin{eqnarray}
\frac{A_{cyl}}{A_{uc}}=\frac{NA_{cyl}}{A_{eff}}.
\end{eqnarray}
For the homogenized system with the rectangular shape considered in this work (6 rows of 10 scatterers with triangular periodicity), the effective thickness $L_{eff}$ follows
\begin{eqnarray}
L_{eff}=5\frac{\sqrt{3}a}{2},
\end{eqnarray}
where $a$ is the lattice constant of the square array of the inner
structure of the homogeneous material. For the parameters
considered in this work, $L_{eff}=$0.549 m.

Finally, the density of the effective medium is considered here as the dynamic mass density,
\begin{eqnarray}
\frac{\rho_{eff}}{\rho_h}=\frac{\rho_h+\rho_s-f(\rho_{h}-\rho_s)}{\rho_h+\rho_s+f(\rho_{h}-\rho_s)},
\end{eqnarray}
where, $\rho_{s}$ is the density of the scatterer and $\rho_{h}$ is the density of the host material.

It would be interesting to know what is the dispersion relation corresponding to this medium with negative elastic modulus.
In the regime of low frequencies, the real ($x=Re(\kappa)$) and
imaginary ($y=-Im(\kappa)$) parts of the bulk modulus can be related
to the propagation constant of the media as \cite{Fang06},
\begin{eqnarray}
\label{eq:dispersion1}
Re(k)=-\frac{\omega}{2}\sqrt{\frac{\rho}{x^2+y^2}}\left(\sqrt{x^2+y^2}-x\right)^{1/2},\\
\label{eq:dispersion2}
Im(k)=\frac{\omega}{2}\sqrt{\frac{\rho}{x^2+y^2}}\left(\sqrt{x^2+y^2}+x\right)^{1/2}.
\end{eqnarray}
In Figure \ref{fig:keff}b, one can observe the dispersion relation
obtained from the effective elastic modulus, $\kappa_{eff}$ in Equation
\ref{eq:effective_modulus}, using Equations \ref{eq:dispersion1} and \ref{eq:dispersion2}. Two spectral bands of no propagating
modes are expected in the vicinity of the resonances of the local
resonators.

From the Fresnel Equation of stratified media, it is possible to calculate
the transmission coefficient of a slab of $0.549$ m of the acoustic
metamaterial analysed here. In Figure \ref{fig:keff}c, we can observe the
absolute value of the transmission coefficient. One can observe a reduction of
the transmission around the resonant frequencies.

In Figure \ref{fig:keff}d, we show data from measurements of the IL of a periodic array of U-profiles distributed in a triangular lattice. We observe that the attenuation peaks predicted by using the effective medium approximation are also experimentally seen. On the other hand, we would like to note that the diffraction limit is near to the second peak, and in the experimental measurements, one can observe Bragg's peak due to the periodicity.
    
    \section{Conclusions}
    \label{sec:conclu}
       In summary, the resonances of a scatterer with complex geometry have been studied from the analysis of the resonances of simpler geometries. The easy way to design the resonances of the scatterers presented in this work opens several possibilities to create an arrangement of resonant scatterers that attenuate a wide range of frequencies below the BG of the structure. The first possibility is to analyse the acoustical behaviour of a periodic array made of several scatterers with different length and cavities. Following the rules for the design of the resonance frequency of both elastic beam and cavity resonances, it is possible to design tunable stop bands in the propagating range of a SC if it were made of rigid scatterers. The results shown in this paper could be used to design effective Sonic Crystal Acoustic Barriers with wide tunable attenuation band in the low-frequency range.
 
    \section{Acknowledgements}
    This work was supported by MEC (Spanish Government) and FEDER
funds, under grants MAT2009-09438 and MTM2009-14483-C02-02. AK and OU are grateful for the support of EPSRC (UK) through research grant EP/E063136/1.

\appendix
\section{End correction formula}
\label{ap1}
The end correction obtained by Norris \textit{et al.} in reference \cite{Norris05} for cylindrical rigid resonators is:
\begin{eqnarray}
\delta \simeq \frac{2a\alpha}{\pi}\log\left(\frac{2}{\alpha}\right),
\end{eqnarray}
where $\alpha$ is the angle of the aperture of the cylindrical Helmholtz resonator and $a$ the radius of the cavity. Taking into account that the resonant frequency of the 2D Helmholtz resonators depend on the area and the aperture of the cavity, we have adapted this formula to our case considering a cylindrical resonator with the same area and aperture as our U-profile. Then the equivalent radius and aperture angle follow
\begin{eqnarray}
a_e=\sqrt{\frac{l_xL_2}{\pi}},\\
\alpha_e=2\arcsin{\left(\frac{L_2}{2a_e}\right)}.
\end{eqnarray}
These definitions need additional conditions that make the aperture angle $\alpha_e$ real. The condition $|\sin(\alpha_e/2)|\leq1$ that gives $L_2 \leq 4 l_x /\pi$. With this, the end correction follows
\begin{eqnarray}
\delta\simeq\frac{2a_e\alpha_e}{\pi}\log{\left(\frac{2}{\alpha_e}\right)}.
\end{eqnarray}


\end{document}